\pdfoutput=1
\documentclass[12pt,a4paper]{article}

\usepackage{ifthen} 
\usepackage{comment}
\newboolean{pdflatex}
\setboolean{pdflatex}{true} 

\newboolean{articletitles}
\setboolean{articletitles}{true} 

\newboolean{uprightparticles}
\setboolean{uprightparticles}{false} 

\usepackage{multirow}
\usepackage{siunitx}
\def\paperauthors{LHCb collaboration} 
\def\paperasciititle{Search for tau- -> mu- mu+ mu- decays at the LHCb experiment with Run 2 data} 
\def\papertitle{Search for \TauToMuMuMu decays at the \lhcb experiment with Run 2 data} 
\def\paperkeywords{{High Energy Physics}, {LHCb}} 
\def\papercopyright{\the\year\ CERN for the benefit of the LHCb collaboration} 
\def\paperlicence{CC BY 4.0 licence}
\def\paperlicenceurl{https://creativecommons.org/licenses/by/4.0/}

\newif\ifEnableSectionTOCLinks
\EnableSectionTOCLinksfalse 

\usepackage[top=1in, bottom=1.25in, left=1in, right=1in]{geometry}

\usepackage{microtype}
\usepackage{lineno}  
\usepackage{xspace} 
\usepackage{caption} 

\usepackage{graphicx}  
\usepackage{color}
\usepackage{colortbl}
\graphicspath{{./figs/}} 

\usepackage{amsmath} 
\usepackage{amssymb}
\usepackage{amsfonts}
\usepackage{upgreek} 

\newcommand*\patchAmsMathEnvironmentForLineno[1]{%
\expandafter\let\csname old#1\expandafter\endcsname\csname #1\endcsname
\expandafter\let\csname oldend#1\expandafter\endcsname\csname
end#1\endcsname
 \renewenvironment{#1}%
   {\linenomath\csname old#1\endcsname}%
   {\csname oldend#1\endcsname\endlinenomath}%
}
\newcommand*\patchBothAmsMathEnvironmentsForLineno[1]{%
  \patchAmsMathEnvironmentForLineno{#1}%
  \patchAmsMathEnvironmentForLineno{#1*}%
}
\AtBeginDocument{%
\patchBothAmsMathEnvironmentsForLineno{equation}%
\patchBothAmsMathEnvironmentsForLineno{align}%
\patchBothAmsMathEnvironmentsForLineno{flalign}%
\patchBothAmsMathEnvironmentsForLineno{alignat}%
\patchBothAmsMathEnvironmentsForLineno{gather}%
\patchBothAmsMathEnvironmentsForLineno{multline}%
\patchBothAmsMathEnvironmentsForLineno{eqnarray}%
}

\usepackage[pdftex,
            pdfauthor={\paperauthors},
            pdftitle={\paperasciititle},
            pdfkeywords={\paperkeywords}]{hyperref}
\usepackage{hyperxmp}
\hypersetup{
    pdfcopyright={Copyright (C) \papercopyright},
    pdflicenseurl={\paperlicenceurl}
}

\usepackage[colorinlistoftodos,textsize=scriptsize]{todonotes}

\usepackage[bottom,flushmargin,hang,multiple]{footmisc}

\usepackage[all]{hypcap} 

\usepackage{xspace} 
\usepackage{upgreek}

\def\lhcb   {\mbox{LHCb}\xspace}

\def\cms    {\mbox{CMS}\xspace}

\def\belletwo {\mbox{Belle~II}\xspace}

\def\MagUp {\mbox{\em Mag\kern -0.05em Up}\xspace}

\ifthenelse{\boolean{uprightparticles}}%
{

 \def\Pmu         {\ensuremath{\upmu}\xspace}

 \def\Ppi         {\ensuremath{\uppi}\xspace}

 \def\Ptau        {\ensuremath{\uptau}\xspace}

 \def\Ppsi        {\ensuremath{\uppsi}\xspace}

 \def\PDelta      {\ensuremath{\Delta}\xspace}                 
 \def\PXi         {\ensuremath{\Xi}\xspace}                 
 \def\PLambda     {\ensuremath{\Lambda}\xspace}                 
 \def\PSigma      {\ensuremath{\Sigma}\xspace}                 
 \def\POmega      {\ensuremath{\Omega}\xspace}                 
 \def\PUpsilon    {\ensuremath{\Upsilon}\xspace}
 \let\oldPi\Pi
 \def\PPi         {\ensuremath{\oldPi}\xspace}

 \def\PB      {\ensuremath{\mathrm{B}}\xspace}                 
 \def\PD      {\ensuremath{\mathrm{D}}\xspace}                 
 \def\PJ      {\ensuremath{\mathrm{J}}\xspace}                 
 \def\PK      {\ensuremath{\mathrm{K}}\xspace}                 
 \def\Pb      {\ensuremath{\mathrm{b}}\xspace}                 
 \def\Pc      {\ensuremath{\mathrm{c}}\xspace}

 \def\Ps      {\ensuremath{\mathrm{s}}\xspace}

 \def\thebaroffset{0.0em}
}
{

 \def\Pmu         {\ensuremath{\mu}\xspace}

 \def\Ppi         {\ensuremath{\pi}\xspace}

 \def\Ptau        {\ensuremath{\tau}\xspace}

 \def\Ppsi        {\ensuremath{\psi}\xspace}                 
                  
 \mathchardef\PDelta="7101
 \mathchardef\PXi="7104
 \mathchardef\PLambda="7103
 \mathchardef\PSigma="7106
 \mathchardef\POmega="710A
 \mathchardef\PUpsilon="7107
 \mathchardef\PPi="7105
 \def\PB      {\ensuremath{B}\xspace}                 
 \def\PD      {\ensuremath{D}\xspace}                 
 \def\PJ      {\ensuremath{J}\xspace}                 
 \def\PK      {\ensuremath{K}\xspace}                 
 \def\Pb      {\ensuremath{b}\xspace}                 
 \def\Pc      {\ensuremath{c}\xspace}

 \def\Ps      {\ensuremath{s}\xspace}

 \def\thebaroffset{0.18em}
}
\newcommand{\offsetoverline}[2][\thebaroffset]{\kern #1\overline{\kern -#1 #2}}%

\makeatletter
\ifcase \@ptsize \relax
  \newcommand{\miniscule}{\@setfontsize\miniscule{4}{5}}
\or
  \newcommand{\miniscule}{\@setfontsize\miniscule{5}{6}}
\or
  \newcommand{\miniscule}{\@setfontsize\miniscule{5}{6}}
\fi
\makeatother

\DeclareRobustCommand{\optbar}[1]{\shortstack{{\miniscule (\rule[.5ex]{1.25em}{.18mm})}
  \\ [-.7ex] $#1$}}

\def\mup        {{\ensuremath{\Pmu^+}}\xspace}
\def\mun        {{\ensuremath{\Pmu^-}}\xspace} 

\def\taum       {{\ensuremath{\Ptau^-}}\xspace}

\def\squark    {{\ensuremath{\Ps}}\xspace}

\def\cquark    {{\ensuremath{\Pc}}\xspace}
\def\cquarkbar {{\ensuremath{\overline \cquark}}\xspace}

\def\bquark    {{\ensuremath{\Pb}}\xspace}
\def\bquarkbar {{\ensuremath{\overline \bquark}}\xspace}

\def\pion   {{\ensuremath{\Ppi}}\xspace}

\def\pip    {{\ensuremath{\pion^+}}\xspace}
\def\pim    {{\ensuremath{\pion^-}}\xspace}

\def\kaon    {{\ensuremath{\PK}}\xspace}

\def\KorKbar {\kern \thebaroffset\optbar{\kern -\thebaroffset \PK}{}\xspace}

\def\Kp      {{\ensuremath{\kaon^+}}\xspace}

\def\D       {{\ensuremath{\PD}}\xspace}

\def\DorDbar {\kern \thebaroffset\optbar{\kern -\thebaroffset \PD}\xspace}

\def\Dp      {{\ensuremath{\D^+}}\xspace}
\def\Dm      {{\ensuremath{\D^-}}\xspace}

\def\DpDm    {\ensuremath{\Dp {\kern -0.16em \Dm}}\xspace}

\def\Dsm     {{\ensuremath{\D^-_\squark}}\xspace}

\def\DmorDsm {{\ensuremath{\D{}_{(\squark)}^-}}\xspace}

\def\B       {{\ensuremath{\PB}}\xspace}

\def\BorBbar {\kern \thebaroffset\optbar{\kern -\thebaroffset \PB}\xspace}

\def\Bd      {{\ensuremath{\B^0}}\xspace}

\def\BdorBdbar {\kern \thebaroffset\optbar{\kern -\thebaroffset \Bd}\xspace}

\def\Bs      {{\ensuremath{\B^0_\squark}}\xspace}

\def\BsorBsbar {\kern \thebaroffset\optbar{\kern -\thebaroffset \Bs}\xspace}

\def\jpsi     {{\ensuremath{{\PJ\mskip -3mu/\mskip -2mu\Ppsi}}}\xspace}

\def\Y#1S{\ensuremath{\PUpsilon{(#1S)}}\xspace}

\def\LorLbar     {\kern \thebaroffset\optbar{\kern -\thebaroffset \PLambda}\xspace}

\def\BF         {{\ensuremath{\mathcal{B}}}\xspace}

\newcommand{\decay}[2]{\ensuremath{\mathinner{#1\!\to #2}}\xspace}

\def\to                 {\ensuremath{\rightarrow}\xspace}

\def\AT#1     {\ensuremath{A_{\mathrm{T}}^{#1}}\xspace}           

\def\C#1      {\ensuremath{\mathcal{C}_{#1}}\xspace}                       
\def\Cp#1     {\ensuremath{\mathcal{C}_{#1}^{'}}\xspace}                    
\def\Ceff#1   {\ensuremath{\mathcal{C}_{#1}^{\mathrm{(eff)}}}\xspace}        
\def\Cpeff#1  {\ensuremath{\mathcal{C}_{#1}^{'\mathrm{(eff)}}}\xspace}       
\def\Ope#1    {\ensuremath{\mathcal{O}_{#1}}\xspace}                       
\def\Opep#1   {\ensuremath{\mathcal{O}_{#1}^{'}}\xspace}                    

\newcommand{\nospaceunit}[1]{\ensuremath{\text{#1}}}       
\newcommand{\aunit}[1]{\ensuremath{\text{\,#1}}}       

\newcommand{\tev}{\aunit{Te\kern -0.1em V}\xspace}
\newcommand{\gev}{\aunit{Ge\kern -0.1em V}\xspace}
\newcommand{\mev}{\aunit{Me\kern -0.1em V}\xspace}
\newcommand{\kev}{\aunit{ke\kern -0.1em V}\xspace}
\newcommand{\ev}{\aunit{e\kern -0.1em V}\xspace}
 
\newcommand{\mevc}{\ensuremath{\aunit{Me\kern -0.1em V\!/}c}\xspace}
\newcommand{\gevc}{\ensuremath{\aunit{Ge\kern -0.1em V\!/}c}\xspace}
\newcommand{\mevcc}{\ensuremath{\aunit{Me\kern -0.1em V\!/}c^2}\xspace}
\newcommand{\gevcc}{\ensuremath{\aunit{Ge\kern -0.1em V\!/}c^2}\xspace}

\def\mum  {\ensuremath{\,\upmu\nospaceunit{m}}\xspace}

\def\fb   {\ensuremath{\aunit{fb}}\xspace}
\def\invfb   {\ensuremath{\fb^{-1}}\xspace}

\def\gsim{{~\raise.15em\hbox{$>$}\kern-.85em
          \lower.35em\hbox{$\sim$}~}\xspace}
\def\lsim{{~\raise.15em\hbox{$<$}\kern-.85em
          \lower.35em\hbox{$\sim$}~}\xspace}

\def\sPlot{\mbox{\em sPlot}\xspace}

\def\sqs   {\ensuremath{\protect\sqrt{s}}\xspace}

\def\pt         {\ensuremath{p_{\mathrm{T}}}\xspace}

\def\ptot       {\ensuremath{p}\xspace}

\def\evtgen     {\mbox{\textsc{EvtGen}}\xspace}

\def\geant      {\mbox{\textsc{Geant4}}\xspace}

\def\photos     {\mbox{\textsc{Photos}}\xspace}

\def\pythia     {\mbox{\textsc{Pythia}}\xspace}

\def\tell1  {TELL1\xspace}
\def\ukl1   {UKL1\xspace}

\newcommand{\ie}{\mbox{\itshape i.e.}\xspace}

\newcommand{\lhcborcid}[1]{\href{https://orcid.org/#1}{\hspace*{0.1em}\raisebox{-0.45ex}{\includegraphics[width=1em]{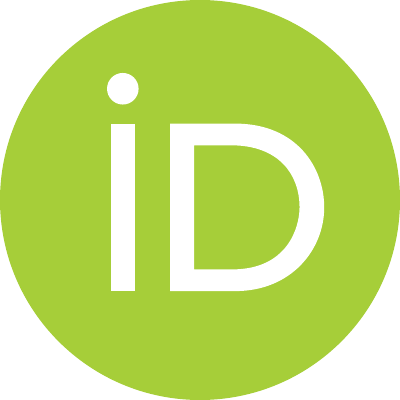}}}}

\hypersetup{
  colorlinks   = true, 
  urlcolor     = blue, 
  linkcolor    = blue, 
  citecolor    = red   
}

\ifEnableSectionTOCLinks
    \usepackage[explicit]{titlesec} 
    
    \let\oldcontentsline\contentsline
    \renewcommand

    \titleformat{\section}{\normalfont\Large\bf}{\hyperlink{tocsection.\thesection}{{\thesection} \parbox[t]{\dimexpr\textwidth-1pc}{#1}}}{1pc}{}

    \titleformat{\subsection}{\normalfont\bf}{\hyperlink{tocsubsection.\thesubsection}{{\thesubsection} \parbox[t]{\dimexpr\textwidth-1pc}{#1}}}{1pc}{}

    \titleformat{name=\section,numberless}[display]{}{}{0pt}{\normalfont\Huge\bfseries #1}
\fi

\usepackage{cite} 
\usepackage{LHCb/mciteplus}

\def\TauToMuMuMu       {\ensuremath{\taum\to\mun\mup\mun}\xspace}

\def\DmToKpipi     {\ensuremath{\Dm\to\pim \Kp\pim}\xspace}

\def\DmTopipipi    {\ensuremath{\Dm\to\pim\pip\pim}\xspace}

\def\DsmTopipipi    {\ensuremath{\Dsm\to\pim\pip\pim}\xspace}

\def\DsToPhiPi     {\ensuremath{\Dsm\to\phi(\to\mun\mup)\pim}\xspace}

\def\DsToetappmunu {\ensuremath{\Dsm\to\eta^{(\prime)}(\to\mun\mup\gamma)\mun\bar{\nu}_{\mu}}\xspace}

\def\DsToTaunu{\ensuremath{\Dsm\to\taum\bar{\nu}_{\tau}}\xspace}

\def\phifull   {{\ensuremath{\phi(1020)}}\xspace}

\def\xgbac{\textit{C}$_{\text{\!AC}}$\xspace}
\def\xgbpid{\textit{C}$_{\text{PID}}$\xspace}

\usepackage{cleveref}

\usepackage{longtable} 
\usepackage{booktabs}

\begin{document}

\renewcommand{\thefootnote}{\fnsymbol{footnote}}
\setcounter{footnote}{1}


\begin{titlepage}
\pagenumbering{roman}

\vspace*{-1.5cm}
\centerline{\large EUROPEAN ORGANIZATION FOR NUCLEAR RESEARCH (CERN)}
\vspace*{1.5cm}
\noindent
\begin{tabular*}{\linewidth}{lc@{\extracolsep{\fill}}r@{\extracolsep{0pt}}}
\ifthenelse{\boolean{pdflatex}}
{\vspace*{-1.5cm}\mbox{\!\!\!\includegraphics[width=.14\textwidth]{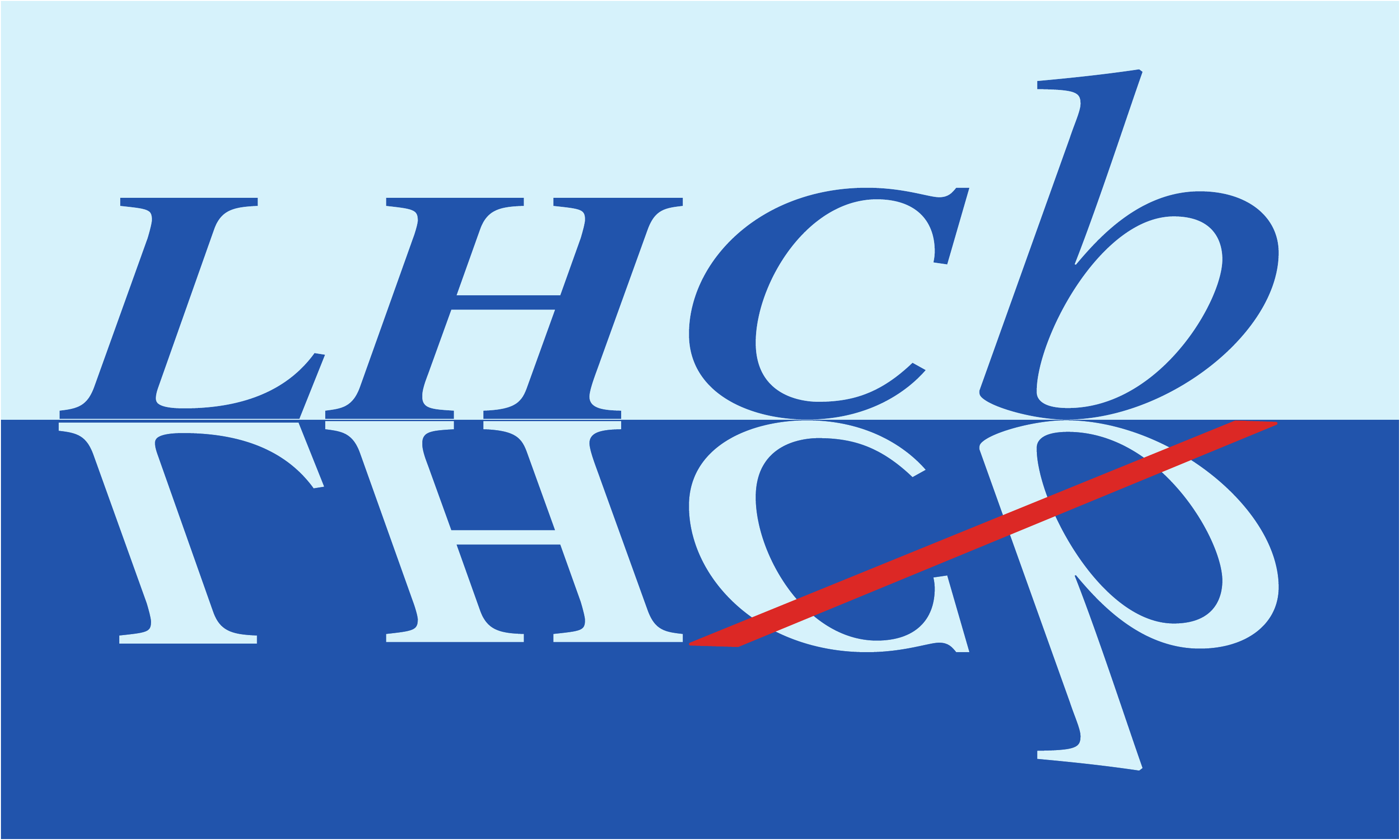}} & &}%
{\vspace*{-1.2cm}\mbox{\!\!\!\includegraphics[width=.12\textwidth]{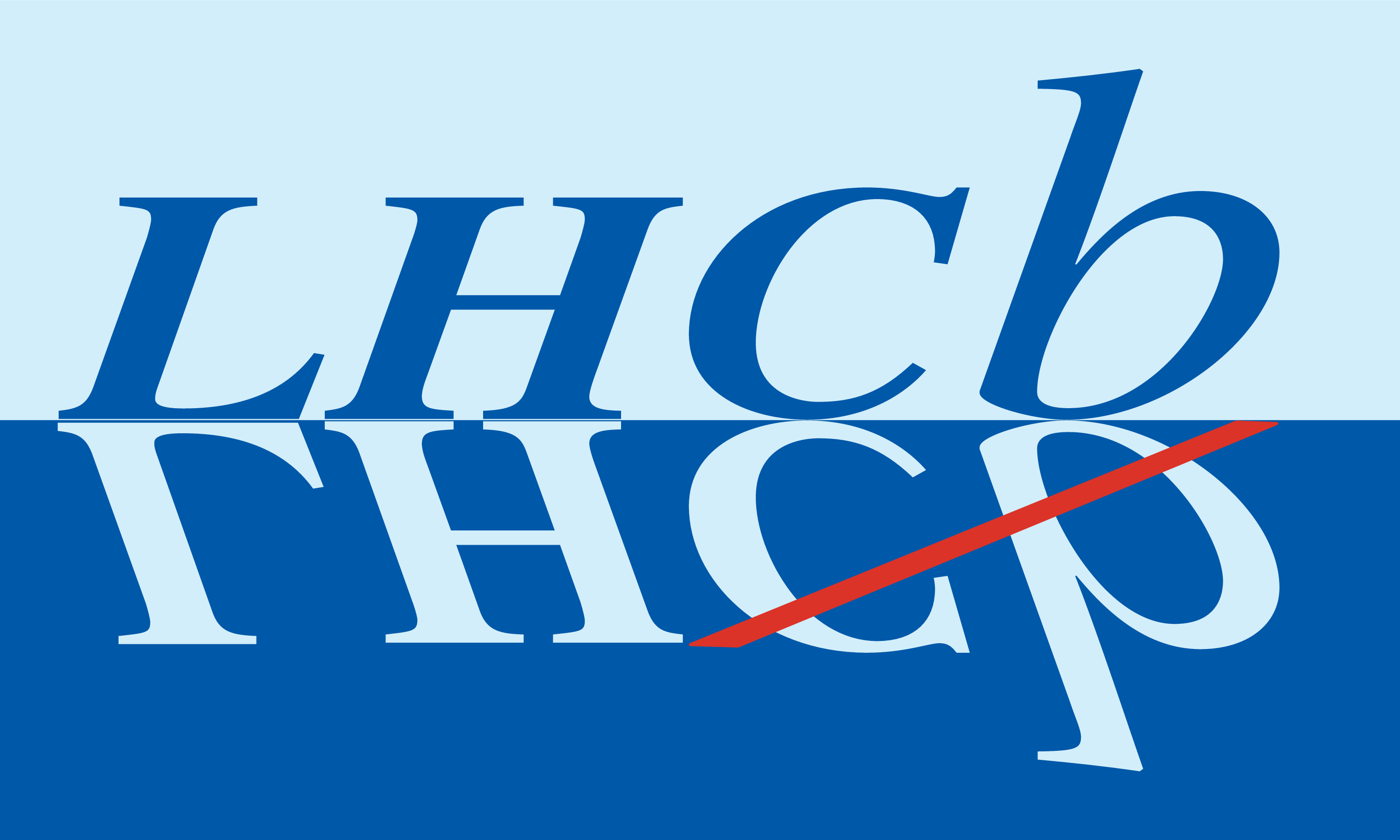}} & &}%
\\
 & & CERN-EP-2026-003 \\  
 & & LHCb-PAPER-2025-052\\  
 & & September 4, 2026 \\ 
 & & \\
\end{tabular*}

\vspace*{4.0cm}

{\normalfont\bfseries\boldmath\huge
\begin{center}
  \papertitle 
\end{center}
}

\vspace*{2.0cm}

\begin{center}
\paperauthors\footnote{Authors are listed at the end of this paper.}
\end{center}

\vspace{\fill}

\begin{abstract}
  \noindent
   A search for the lepton-flavour-violating decay \TauToMuMuMu is carried out using data collected by the \lhcb experiment between 2016 and 2018 in proton-proton collisions at the LHC at a centre-of-mass energy of $13\tev$, corresponding to an integrated luminosity of 5.4\invfb. An upper limit of $1.9\,(2.3)\times 10^{-8}$ is set at the 90\% (95\%) confidence level on the branching fraction of the decay. 
  
\end{abstract}

\vspace*{2.0cm}

\begin{center}
  Published in Journal of High Energy Physics 09 (2026) 069
\end{center}

\vspace{\fill}

{\footnotesize 
\centerline{\copyright~\papercopyright. \href{\paperlicenceurl}{\paperlicence}.}}
\vspace*{2mm}

\end{titlepage}


\newpage
\setcounter{page}{2}
\mbox{~}
%
%
%
%


\renewcommand{\thefootnote}{\arabic{footnote}}
\setcounter{footnote}{0}


\cleardoublepage


\pagestyle{plain} 
\setcounter{page}{1}
\pagenumbering{arabic}


\section{Introduction}
\label{sec:Introduction}

Lepton-flavour-violating decays are forbidden in the Standard Model~(SM) under the assumption of massless neutrinos. In a massive-neutrino scenario, lepton-flavour-violating decays are expected to occur at extremely low rates, with branching fractions around $10^{-55}$~\cite{Blackstone:2019njl}, well below the sensitivity of any current or foreseen experimental facilities. Nevertheless, many extensions of the SM including heavy neutrinos~\cite{Cvetic:2002jy, Akeroyd:2006bb} or the exchange of an extra neutral gauge boson $Z^\prime$~\cite{Yue:2002ja} predict much larger branching fractions at the level of \mbox{$10^{-10}$--$10^{-8}$}. Therefore, observation of the \TauToMuMuMu decay\footnote{Charge conjugation is implied
throughout the paper.} would indicate new physics beyond the SM, while further improving the experimental limit on its branching fraction would provide constraints on SM extensions.

A previous search with the LHCb experiment using the Run~1 (2011--2012) dataset set an upper limit on the branching fraction of the \TauToMuMuMu decay of \mbox{$\mathcal{B}(\TauToMuMuMu) <4.6\,(5.6)\times10^{-8}$} at the 90\% (95\%) confidence level (CL)~\cite{LHCb-PAPER-2014-052}. The ATLAS collaboration also conducted a search for the \TauToMuMuMu decay using data collected in 2012, obtaining an upper limit of $3.8\times10^{-7}$ at 90\% CL~\cite{ATLAS:2016jts}. A more recent search by the \cms collaboration, based on the full Run~2 dataset, established an upper limit of $2.9\,(3.6)\times 10^{-8}$ at 90\% (95\%) CL~\cite{CMS:2023iqy}. The most stringent upper limit from a single experiment was reported by the \belletwo collaboration, which obtained $1.9 \times 10^{-8}$ at 90\% CL~\cite{Belle-II:2024sce}.

In this paper, a search for the lepton-flavour-violating \TauToMuMuMu decay is presented, based on data collected by the LHCb experiment in proton-proton ($pp$) collisions during the years 2016--2018 of Run~2 of the LHC at a centre-of-mass energy of $\sqs=13\tev$, corresponding to an integrated luminosity of 5.4\invfb. 
The \TauToMuMuMu branching fraction is evaluated relative to the known \decay{\Dsm}{\phifull \pim} branching fraction, accounting for the fraction of \taum leptons produced in \Dsm decays.\footnote{In the following, $\phi$ denotes the \phifull meson.} 
This normalisation channel, where the intermediate state decays as \decay{\phi}{\mun\mup}, is chosen for its similar topology and decay kinematics with respect to the signal.
The \TauToMuMuMu branching fraction is given by
\begin{equation}
\begin{aligned}
\BF(\TauToMuMuMu) & = \BF(\DsToPhiPi) \times \frac{f_{D_s}^{\tau}}{\BF(\DsToTaunu)} \times \frac{\epsilon_{D_s}}{\epsilon_{\tau}} \times \frac{N_{\tau}}{N_{D_s}} \\
      &  = \alpha \times N_{\tau}\,,
\end{aligned}
\label{BF}
\end{equation}
where $\BF(\DsToPhiPi)$ and $\BF(\DsToTaunu)$ are taken from Ref.~\cite{PDG2024}, $f_{D_s}^{\tau}$ is the fraction of \taum leptons that are produced by \Dsm  decays versus the inclusive production from heavy-hadron decays, 
$\epsilon_{\tau}$ ($\epsilon_{D_s}$) is the efficiency for selecting candidates in the signal~(normalisation) channel, and $N_{\tau}$ ($N_{D_s}$)  the number of candidates observed in the signal~(normalisation) channel. The total normalisation factor, $\alpha$, is referred to as the \textit{single-event sensitivity}. 

This article is organised as follows: in section~\ref{sec:DetectorAndDataset},
the LHCb detector, data and simulation samples used in the analysis are described; 
the selection criteria for the signal and normalisation channels
are outlined in section~\ref{sec:Selection}; 
the corrections applied to the simulation to improve the description of the data are reported in section~\ref{sec:CorrectionToSimulation}; 
the treatment of the different sources of background is discussed in section~\ref{sec:Background}; 
section~\ref{sec:Normalisation}
reports on the evaluation of the selection efficiencies, the normalisation factors, and 
the corresponding systematic uncertainties; 
the procedure to measure the signal yield is discussed in section~\ref{sec:MassFit}; section~\ref{sec:Results} reports the strategy adopted for the evaluation of the upper limit and the corresponding results;  
and finally, conclusions are drawn in section~\ref{sec:Conclusion}.

\section{Detector and dataset}
\label{sec:DetectorAndDataset}

The \lhcb detector~\cite{LHCb-DP-2008-001,LHCb-DP-2014-002} is a single-arm forward
spectrometer covering the \mbox{pseudorapidity} range $2<\eta <5$,
designed for the study of particles containing \bquark or \cquark
quarks. The detector used to collect data for this analysis includes a high-precision tracking system
consisting of a silicon-strip vertex detector surrounding the $pp$
interaction region~\cite{LHCb-DP-2014-001}, a large-area silicon-strip detector located
upstream of a dipole magnet with a bending power of about
$4{\mathrm{\,T\,m}}$, and three stations of silicon-strip detectors and straw
drift tubes~\cite{LHCb-DP-2017-001} placed downstream of the magnet.
The tracking system provides a measurement of the momentum, \ptot, of charged particles with
a relative uncertainty that varies from 0.5\% at low momentum to 1.0\% at 200\gevc.
The minimum distance of a track to a primary $pp$ collision vertex (PV), the impact parameter (IP), 
is measured with a resolution of $(15+29/\pt)\mum$,
where \pt is the component of the momentum transverse to the beam, in\,\gevc.
Different types of charged hadrons are distinguished using information
from two ring-imaging Cherenkov detectors~\cite{LHCb-DP-2012-003}. 
Photons, electrons and hadrons are identified by a calorimeter system consisting of
scintillating-pad and preshower detectors, an electromagnetic and a hadronic calorimeter. Muons are identified by a
system composed of alternating layers of iron and multiwire
proportional chambers~\cite{LHCb-DP-2012-002}.

The online event selection is performed by a trigger~\cite{LHCb-DP-2012-004}, 
which consists of a hardware stage, based on information from the calorimeter and muon
systems, followed by a software stage, which applies a full event
reconstruction. The hardware trigger selects candidates with either one or two muons with high \pt, while the software trigger adds requirements on the kinematics, track quality and particle-identification (PID) information of the muons. The trigger requirements at the second stage of the software trigger (HLT2) differ between the signal and the normalisation channel, as a dedicated trigger selection for the three-muon final state was available for the signal. In addition, the configuration used to select \DsToPhiPi candidates in the 2016 data sample differs from that employed in 2017 and 2018 at the HLT2 level, where additional trigger lines were available, resulting in a lower selection efficiency in 2016 compared to the other years.
Triggered data further undergo a centralised, offline processing step to deliver physics-analysis-ready data across the entire \lhcb physics programme~\cite{Stripping}.

Simulated samples of \TauToMuMuMu and \DsToPhiPi decays are produced to understand the effects of detector acceptance and the imposed selection requirements. In the simulation, $pp$ collisions are generated using \pythia~\cite{Sjostrand:2007gs} with a specific LHCb configuration~\cite{LHCb-PROC-2010-056}. Decays of unstable particles are
described by \evtgen~\cite{Lange:2001uf}, in which final-state radiation is generated using \photos~\cite{davidson2015photos}.
The interaction of the generated particles with the detector, and its response, are implemented using the \geant toolkit~\cite{Agostinelli:2002hh} as described in Ref.~\cite{LHCb-PROC-2011-006} and further corrected to account for known differences between data and simulation in kinematic distributions, detector response, and particle-identification variables of the reconstructed candidates.
The \DsToPhiPi mode is exploited to derive these corrections based on the similarity observed in the distributions of the aforementioned variables with respect to the \TauToMuMuMu channel, as described in section~\ref{sec:CorrectionToSimulation}.
Samples are produced accounting for the main sources of \taum leptons in $pp$ collision. The \taum production cross-section is dominated by \DmorDsm mesons produced either promptly or in \bquark-hadron~(secondary) decays, as well as by direct decays of \bquark hadrons.
Contributions from $Z$ and $W$ boson decays as well as from the Drell--Yan process are orders of magnitude smaller than that from heavy-hadron production and are therefore neglected~\cite{LHCb-PAPER-2015-049, LHCb-PAPER-2018-016}. 
The \TauToMuMuMu decay is generated with a uniform phase-space distribution.

\section{Selection}
\label{sec:Selection}
The signal \TauToMuMuMu decay and the \DsToPhiPi decay used as the normalisation channel are selected applying criteria that are as similar as possible. The \taum and \Dsm candidates are reconstructed by combining three tracks with a total charge of $-1$. 
The three tracks are required to be compatible with originating from a common vertex and have significant $\chi^2_{\rm IP}$ with respect to all PVs. 
Here, $\chi^2_{\rm IP}$ is defined as the difference in the vertex-fit $\chi^2$ of a given PV reconstructed with and without the track under consideration. 
Loose requirements on the track quality and \pt are placed on all tracks. 
In case of \Dsm candidates, two oppositely charged tracks should be identified as muons, with the same muon-identification requirement as applied to all three final-state tracks in the \taum case.
Muon identification is based on the track momentum and the pattern of hits in the muon chambers associated with the track~\cite{LHCb-DP-2013-001}.
In the selection of \Dsm candidates,
the \mun\mup\pim mass is required to lie within $\pm50\mevcc$ of the known \Dsm mass~\cite{PDG2024}, while the \mun\mup mass must fall within $\pm20\mevcc$ of the known $\phi$-meson mass~\cite{PDG2024}. 
Tau candidates are removed if any of the two \mun\mup combinations lie within the $\pm 30\mevcc$ region around the $\phi$ mass, and are required to have a mass larger than $450\mevcc$ to suppress \DsToetappmunu decays, as discussed in section~\ref{sec:Background}.
The same-sign dimuon mass is instead required to be larger than 250\mevcc in order to reject candidates containing duplicate tracks reconstructed from hits originated by a single charged particle.

The signal sample is divided based on the three-muon mass $M_{\mu\mu\mu}$ into: a signal region, where the signal selection efficiency, and the signal and background yields are evaluated, defined as \mbox{$\vert M_{\mu\mu\mu} - m_\tau \vert \leq 20\mevcc$}, with $m_\tau$ being the known $\tau$ mass~\cite{PDG2024}; and a sideband region with \mbox{$|M_{\mu\mu\mu}-m_\tau|> 20\mevcc$}, which is used to model the background shape and obtain an expectation of the background yield in the signal region. The signal region is masked in data until the end of the analysis, to avoid experimenter's bias.

The main discrimination of \TauToMuMuMu signal over background is achieved by
means of two gradient-boosted decision-tree classifiers, both based on the binary \textsc{XGBoost}~\cite{XGBoosting} model. The first classifier (\xgbac, for \textit{anticombinatorial}), based on topological and kinematic information, is designed to suppress the contribution from random combinations of tracks wrongly associated with the same vertex, referred to as combinatorial background. The second classifier~(\xgbpid), based on PID and kinematic information, is trained to reduce physics background from hadrons misidentified as muons. The main contribution to this background comes from hadronic \DmorDsm decays. 
Both classifiers are trained using simulated signal candidates as a signal proxy and data candidates in the part of the sideband region close to the signal region (\mbox{$20 < |M_{\mu\mu\mu}-m_\tau|\leq 30\mevcc$}) as a background proxy. 
The use of two separate classifiers improves performance by exploiting different features of the decay. 
Supervised learning with \textit{k}-fold ($k=5$) cross-validation \cite{Geisser1993} is used to achieve an unbiased classifier response. 

The largest discrimination power in the \xgbac classifier comes from 
isolation variables that
quantify how much the other tracks of the event are likely to
originate from the same decay as the signal tracks.
The other input variables to the \xgbac classifier are: the \pt and $\chi^2_{\rm IP}$ of each final-state track and of the \taum particle with respect to their own PV; the \taum decay time; the track fit $\chi^2$ for the three muons and the fit $\chi^2$ of their vertex, and the displacement of the latter with respect to the PV; the cosine of the angle between the \taum momentum vector and the vector from the primary to the decay vertex; the number of tracks in a cone built around each final-state track; and the asymmetry between the \pt of the final-state track and the sum of the \pt of all the particles within the cone. 
Inputs to the \xgbpid classifier are the muon- and kaon-identification probabilities and pseudorapidity of each final-state track. The correlation between the two classifiers is found to be negligible. 
Selected \taum candidates are retained if they lie within \xgbac~$>0.80$ and \xgbpid~$>0.88$, corresponding to a signal efficiency of about 60\%. 
An additional requirement is introduced to remove residual hadronic background as discussed in section~\ref{sec:Background}.
No selection based on the \xgbac or \xgbpid classifiers is applied to the normalisation channel, owing to its already high signal purity.

It is also possible to reconstruct decays in which one of the muons falls outside the acceptance of the muon system or has insufficient momentum to penetrate it, by considering candidates for which only two tracks satisfy the muon-identification requirements of the analysis. However, the larger background contamination, together with the limitations of the hardware trigger, which primarily selects tracks with high \pt, make the sensitivity to the \TauToMuMuMu channel significantly worse than in the case where all tracks are identified as muons. The contribution of this category to the upper limit is found to be negligible and is therefore excluded from the final result presented in this paper. 

\section{Simulation calibration}
\label{sec:CorrectionToSimulation}
\begin{figure}
    \centering
    \includegraphics[width=0.6\linewidth]{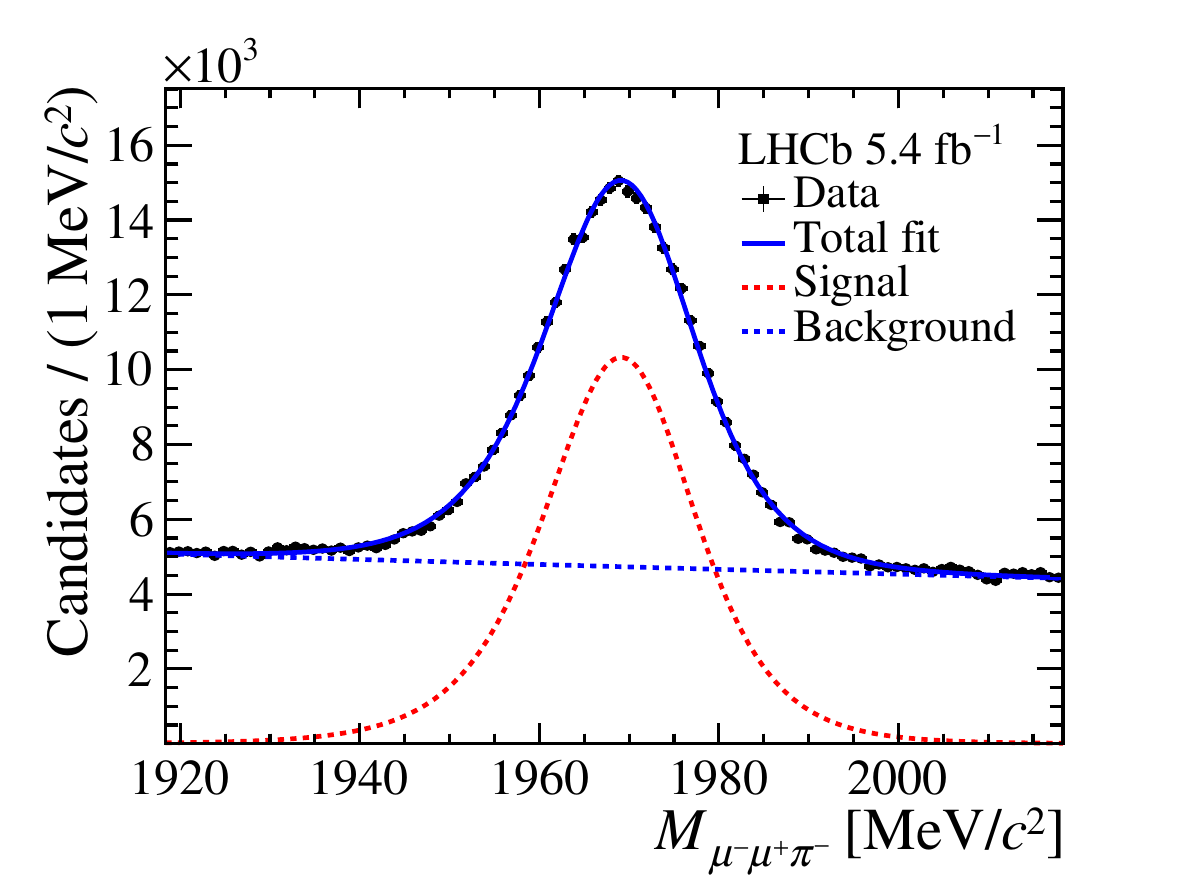}
    \caption{Mass distribution of selected \ensuremath{\mun\mup\pim} candidates together with the fit result.}
    \label{fig:fit_Ds2PhiPi}
\end{figure}

Simulation samples produced for the relevant sources of \taum leptons are combined with relative fractions determined using the \bquark\bquarkbar and \cquark\cquarkbar cross-sections as measured by the 
\lhcb experiment in $pp$ collisions at $\sqs=13\tev$ \cite{{LHCb-PAPER-2015-041, LHCb-PAPER-2015-037}} and the inclusive \decay{\bquark}{\DmorDsm}, \decay{\cquark}{\DmorDsm}, \decay{\bquark}{\taum} and \decay{\DmorDsm}{\taum} branching fractions from Ref.~\cite{PDG2024}. 
The ratio between prompt and secondary charm production is then adjusted to data using the \DsToPhiPi channel by comparing the distributions of the logarithm of $\chi^2_{\rm IP}$ of the \Dsm meson in simulation and background-subtracted data. The background-subtracted distribution is obtained from a maximum-likelihood fit to the \ensuremath{\mun\mup\pim} mass distribution of selected \Dsm candidates, using the \sPlot technique~\cite{Pivk:2004ty}.
The signal probability density function~(PDF) is parametrised by a Johnson's $S_U$ function~\cite{johnsonsu} and the background by an exponential function. The tail parameters of the Johnson's $S_U$ function are fixed from a fit to simulation, while the peak position and width, as well as the slope of the exponential component, are free to vary. 
The fit result is shown in figure~\ref{fig:fit_Ds2PhiPi} for the combined 2016--2018 dataset. 

A weighting algorithm from the \texttt{hep\_ml} package~\cite{Rogozhnikov:2016bdp} based on decision trees is used to correct for global differences in kinematics and detector response. The input features are topological and kinematic variables of the \Dsm and the total number of tracks in the event.
Corrections to the muon-identification variables are evaluated from the ratio of the distributions in data and simulation in different regions of total number of tracks, momentum and pseudorapidity of the muon with the lowest muon-identification probability.
Small residual disagreements between data and simulation are observed in the output of the \xgbac and \xgbpid classifiers when applied to the \DsToPhiPi channel. The ratios of the corresponding one-dimensional distributions in data and simulation are used to derive correction factors that account for these differences. The ratio of the signal- to normalisation-channel efficiencies is further corrected to account for differences between data and simulation arising from track reconstruction and trigger selections in the two channels.
The tracking efficiency is calibrated through a sample of prompt \mbox{\decay{\jpsi}{\mup \mun}} decays~\cite{LHCb-DP-2013-002}, by comparing the performance of the track-finding algorithm obtained excluding at least one subdetector with that obtained including all subdetectors. 
This calibration is determined as a function of the event multiplicity, track $p$ and $\eta$ by computing the overall correction as the product of the single-track efficiency for each final-state particle. The trigger efficiency corrections are derived from a sample of \DsToPhiPi candidates triggered independently of the signal decay chain and comparing the performance of the signal trigger lines in data and simulation in bins of the \pt and $\chi^2_{\rm IP}$ of the final-state track with the highest \pt~\cite{LHCb-DP-2012-004}. The correction factors are evaluated independently for each year of data taking and are of the order (0.1--3.0)\%.

\section{Background}
\label{sec:Background}
Background contamination to the \TauToMuMuMu decay mainly arises from random combinations of muons or misidentified hadrons from decays of heavy or light hadrons, forming a three-track vertex displaced from the PV, with a mass close to that of the \taum lepton. The dominant background processes are charm meson decays into three charged hadrons, all misidentified as muons. These decays are strongly suppressed by the selection based on the \xgbac and \xgbpid classifiers. However, when the pion and kaon mass hypotheses are assigned to the three final-state tracks in all possible combinations within the selected data sample, peaking structures are observed in the mass distributions corresponding to the \pim\Kp\pim and \pim\pip\pim mass hypotheses. These structures are consistent with contributions from \DmToKpipi and \decay{\DmorDsm}{\pim\pip\pim} decays, characterised by branching fractions at the permille to percent level~\cite{PDG2024}. To further suppress the \DmToKpipi contribution, a tight kaon-identification requirement is imposed on the final-state track with charge opposite to that of the parent particle.
This requirement is applied when the \pim\Kp\pim mass hypothesis has a combined mass below $1900\mevcc$ and rejects more than 99.9\% of the \Dm candidates according to simulation. The expected remaining contribution is found to be negligible and is therefore ignored in the fit to the $\mu^-\mu^+\mu^-$ mass distribution. 
In contrast, contaminations from \decay{\DmorDsm}{\pim\pip\pim} decays are not significantly reduced by the application of tighter particle-identification requirements and are therefore modelled in the final fit.
The expected contamination is estimated by fitting the three-track mass distribution in the sidebands and extrapolating the yields of the two sources into the signal region.
The fit is performed twice, once with the \xgbac cut only and once with the \xgbac cut and a reversed \xgbpid cut. 
The difference between the results of the two fits is taken as the estimate of the yield of remaining hadronic decays in the signal region. Contributions from $\decay{\DmorDsm}{V(\to\mun\mup) \pim}$ decays with $V=\eta,\,\rho/\omega,\,\eta^\prime$ are assumed to be subdominant with respect to \decay{\DmorDsm}{\pim\pip\pim} decays and are therefore ignored in the analysis. 

Another source of background comes from decays into three genuine muons, with \DsToetappmunu providing the main contribution. 
Although these decays are largely suppressed by the selection on the minimum dimuon mass, a residual contribution remains in the low-mass sideband, where it exhibits a continuous, decreasing distribution.
This contribution is explicitly taken into account in the fit to the \mun\mup\mun mass distribution in data to correctly estimate the shape of the background in the signal region.

\section{Normalisation and systematic uncertainties}
\label{sec:Normalisation}
The \TauToMuMuMu yield, extracted as described in section~\ref{sec:MassFit}, is converted into a branching fraction through normalising to the yield of \DsToPhiPi decays according to Eq.~(\ref{BF}).
The fraction of $\tau^-$ leptons produced via \Dsm decays, $f^\tau_{D_s}$, is given by the ratio between the production cross-section of \taum leptons from \Dsm decays and the inclusive production cross-section from heavy-hadron decays. 
It is calculated using the \bquark\bquarkbar and \cquark\cquarkbar cross-sections and the inclusive $\decay{\bquark}{\DmorDsm}$, $\decay{\cquark}{\DmorDsm}$, $\decay{\bquark}{\taum}$ and $\decay{\DmorDsm}{\taum}$ branching fractions as defined in section~\ref{sec:CorrectionToSimulation}. 
The efficiency for selecting signal and \DsToPhiPi candidates includes the detector acceptance, reconstruction, particle identification, trigger, and selection effects. It is calculated from simulation after correcting for differences with respect to data, as described in section~\ref{sec:CorrectionToSimulation}. 

The statistical uncertainty on the corrections is considered as a systematic uncertainty in the upper-limit evaluation. Systematic uncertainties on the track reconstruction and trigger efficiency corrections discussed in section~\ref{sec:CorrectionToSimulation} take into account variations in the binning scheme, defined in the relevant kinematic variables for each correction, and, for the track reconstruction efficiency, differences in the hadronic interaction of pions and muons in the detector material. A correction of $(1.8 \pm 0.9)$\% is applied to the normalisation channel to account for the truncated $\phi$ mass window present in simulation. 

The yields of \DsToPhiPi candidates in data are determined from fits to the reconstructed \ensuremath{\mun\mup\pim} mass distributions, as described in section~\ref{sec:CorrectionToSimulation}, performed separately on each data-taking year. To assign a systematic uncertainty on the normalisation yield, the fit to data is repeated using two alternative signal models: a two-sided Crystal Ball function~\cite{Skwarnicki:1986xj} and a two-sided version of the Hypatia function~\cite{Santos:2013gra}. In both cases, the peak position and resolution parameters of the PDFs are free to vary, while the other parameters are fixed from a fit to simulation. The greatest variation obtained on the normalisation yield is taken as its systematic uncertainty. 

Input values to the normalisation factor are reported in table~\ref{tab:normalisation:alpha_summary}, together with the resulting single-event sensitivity $\alpha$ obtained for each data-taking year. The dominant contributions to the systematic uncertainties on $\alpha$ come from external inputs, \ie from the knowledge of the $D_s^-$ decay branching fractions and production fraction $f_{D_s}^\tau$, whereas the leading source of statistical uncertainty on $\alpha$ comes from the determination of the ratio of efficiencies. The \DsToPhiPi yield $N_{D_s}$ and the efficiency ratio $\epsilon_{D_s}/\epsilon_{\tau}$ are lower in the 2016 data sample with respect to the other years due to the less efficient trigger selection described in section~\ref{sec:DetectorAndDataset}.
In the evaluation of the upper limit, the statistical and systematic uncertainties on $\alpha$ are summed in quadrature and treated as a single uncertainty.

\begin{table}[t]
\centering
\caption{Terms entering the normalisation factor and resulting single-event sensitivity $\alpha$. For each term, the statistical and systematic uncertainties are combined and quoted as a single uncertainty which is considered systematic. Branching fractions are taken from Ref.~\cite{PDG2024}. The individual integrated luminosities are 1.6, 1.7, and 2.1~\invfb for 2016, 2017, and 2018, respectively.}
\resizebox{\textwidth}{!}{
\begin{tabular}{l|r@{\:±\:}lr@{\:±\:}lr@{\:±\:}l}
\toprule
& \multicolumn{6}{c}{\textbf{External inputs}} \\
\midrule
$\mathcal{B}(\DsToPhiPi)$ & \multicolumn{2}{c}{} & (1.289 & 0.094) $\times 10^{-5}$  &  \multicolumn{2}{c}{}     \\
$f_{D_s}^{\tau}$          & \multicolumn{2}{c}{} & (78.4  & 3.7) $\times 10^{-2}$  &  \multicolumn{2}{c}{}     \\
$\mathcal{B}(\DsToTaunu)$ & \multicolumn{2}{c}{} & (5.32  & 0.11) $\times 10^{-2}$  &  \multicolumn{2}{c}{}     \\
\midrule
                                  & \multicolumn{2}{c}{\textbf{2016}}   & \multicolumn{2}{c}{\textbf{2017}}   & \multicolumn{2}{c}{\textbf{2018}}   \\
\midrule
$N_{D_s}$                         & (43.1  & 0.6) $\times10^3$             & (88.5   &  1.2) $\times10^3$       & (103.8  & 1.1) $\times 10^3$        \\
$\epsilon_{D_s}/\epsilon_{\tau}$  & 0.322  & 0.022  & 0.562   & 0.034 & 0.575   & 0.036 \\
\midrule
$\alpha$                          & (1.42  & 0.16) $\times10^{-9}$          & (1.21   & 0.13) $\times10^{-9}$     & (1.05   & 0.11) $\times 10^{-9}$     \\
\bottomrule
\end{tabular}
}
\label{tab:normalisation:alpha_summary}
\end{table}

\section{Fit strategy}
\label{sec:MassFit}

In order to increase the sensitivity to the signal decay, the upper limit on the branching fraction is evaluated in bins of the output of the two classifiers \xgbac and \xgbpid. 
The optimal binning scheme is chosen to maximise the signal-to-background separation quantified by the difference between the median expected likelihood ratios under the signal-plus-background and background-only hypotheses following the procedure of Ref.~\cite{LHCb-PAPER-2012-007,Perrin-Terrin:2013cmm}. 
In the optimisation, the signal yield in the full sample is assumed to be unity, and the efficiency in each bin is determined from simulation, while the expected background yield is estimated by fitting the data sidebands with an exponential function and extrapolating the yield into the signal region.
The binning is further constrained to keep the total number of bins commensurate with the size of the available data sample and results in a $4 \times 4$ scheme.

In the final fit to the data, the signal PDF is modelled with a Johnson's $S_U$ function, where the peak position and width parameters are extracted from simulation and calibrated using the normalisation channel, exploiting the similarity of the mass distributions. The background PDF is the sum of three contributions representing the combinatorial, the three-muon and the three-hadron components. The combinatorial background component is modelled by an exponential function, with both the slope and yield parameters free to vary independently in each bin. For the other two background contributions, the mass shapes are assumed to be identical across all bins, as no significant variations are observed in simulated candidates passing the signal selection criteria. The total yields are constrained with a Gaussian prior to the values obtained from a fit to the full data sample. The shape and per-bin fractions of the \DsToetappmunu component are determined from simulation, with the latter constrained in the fit with a Gaussian prior.
The charm background is composed of residual \DmTopipipi and \DsmTopipipi decays. 
It is modelled with two Johnson’s $S_U$ functions, where the peak positions and widths are fixed to the values determined from a fit to the full data sample, while the tail parameters are allowed to vary with a Gaussian constraint with mean and width determined from simulation. 
The per-bin
fractions are extracted from dedicated fits to data in bins of \xgbac or \xgbpid and Gaussian-constrained in the fit.
The relative yield of the \Dm meson with respect to the \Dsm component is fixed according to the value derived from the known branching fractions~\cite{PDG2024}, production rates~\cite{LHCb-PAPER-2015-041} and the corresponding selection efficiencies. 
The fit stability is found to improve when rejecting the least sensitive bin, \ie that corresponding to the lowest values of the \xgbac and \xgbpid outputs, where more background is present. 

For each bin, the expected background yield for the combined 2016--2018 dataset is reported in table~\ref{tab:eventsall}. It is estimated by fitting the data sidebands with the background-only PDF and extrapolating the resulting yield into the signal region.

\begin{table}[t]
\centering
\caption{Number of expected background and observed candidates in the signal region in the 15 \xgbac and \xgbpid bins for the whole dataset integrated over the three data-taking years. The least sensitive bin is excluded from the limit evaluation.}
\label{tab:eventsall}
\begin{tabular}{ccr@{\:±\:}lc}
\toprule
\xgbac bin           & \xgbpid bin & \multicolumn{2}{c}{Expected}  & {Observed} \\ \midrule
\multirow{3}{*}{[0.800, 0.930]} & [0.951, 0.978] & 288.3 & 7.3 &   284 \\
                                    & [0.978, 0.991] & 187.3 & 5.8 &   174 \\
                                    & [0.991, 1.000] & 111.2 & 4.6 &   110 \\
\midrule
\multirow{4}{*}{[0.930, 0.974]} & [0.880, 0.959] & 194.5 & 6.4 &   203 \\
                                    & [0.959, 0.982] & 110.3 & 4.7 &   121 \\
                                    & [0.982, 0.992] &  79.1 & 3.7 &    67 \\
                                    & [0.992, 1.000] &  56.5 & 3.3 &    41 \\
\midrule
\multirow{4}{*}{[0.974, 0.990]} & [0.880, 0.966] & 129.9 & 4.9 &   133 \\
                                    & [0.966, 0.985] &  63.1 & 3.5 &    62 \\
                                    & [0.985, 0.993] &  40.8 & 2.9 &    50 \\
                                    & [0.993, 1.000] &  31.3 & 2.5 &    42 \\
\midrule
\multirow{4}{*}{[0.990, 1.000]} & [0.880, 0.970] &  64.6 & 3.8 &    79 \\
                                    & [0.970, 0.987] &  34.0 & 2.9 &    33 \\
                                    & [0.987, 0.994] &  33.3 & 2.8 &    29 \\
                                    & [0.994, 1.000] &  20.3 & 2.3 &    23 \\ 
\midrule
\multicolumn{2}{l}{Total}                            & 1444  &  17 &  1451 \\ \bottomrule
\end{tabular}
\end{table}

Systematic uncertainties are incorporated into the upper-limit evaluation by treating the single-event sensitivities, the bin-by-bin efficiencies, and the parameters describing the signal and background shapes as nuisance parameters, constrained by Gaussian functions with widths set to their estimated uncertainties.

\section{Results}
\label{sec:Results}

The evaluation of the upper limit on the branching fraction is based on the $\text{CL}_\textit{s}$ method~\cite{Cowan:2010js} and implemented using the \textsc{RooStats} toolkit~\cite{Moneta2010}.
The expected upper limit for the 2016--2018 dataset is found to be $2.0\,(2.5) \times 10^{-8}$ at 90\% (95\%) CL. The impact of the systematic uncertainties is of order a few percent. 

\begin{figure}
    \centering
    \includegraphics[width=0.5\linewidth]{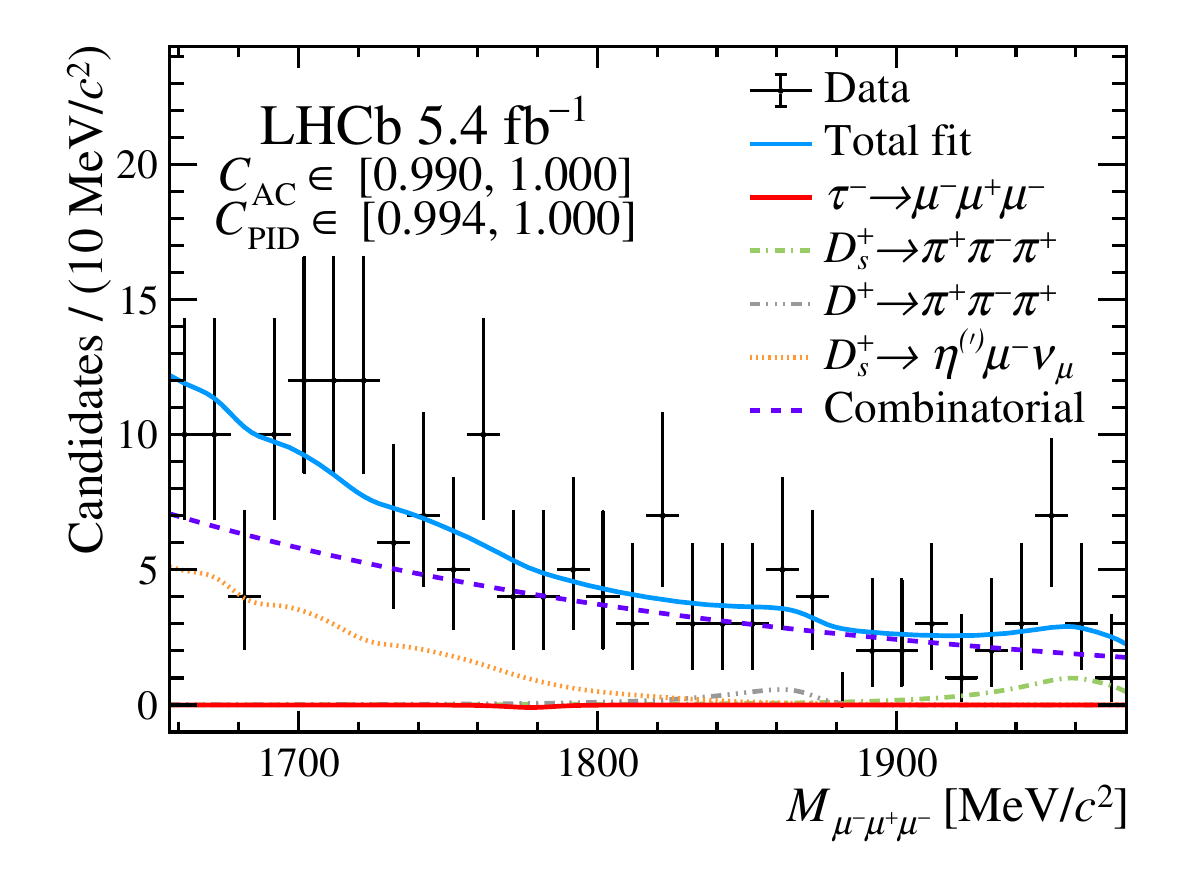}
    \caption{Result of the fit to the mass spectrum of the three muons in the most sensitive bin for the combined dataset.} 
    \label{fig:fit_mostsensitivebin_all}
\end{figure}

The number of candidates observed in each bin in the signal region is reported in the last column of table~\ref{tab:eventsall} for the combined dataset.
An unbinned extended maximum-likelihood fit to the mass spectrum of the three muons is performed simultaneously in the 15 \xgbac and \xgbpid bins for each data-taking year. The total fit PDF is a combination of the background PDF and the signal PDF, discussed in section~\ref{sec:MassFit}. 
The result of the fit in the bin with the largest score in \xgbac and \xgbpid output is shown in figure~\ref{fig:fit_mostsensitivebin_all} for the combined dataset.

\begin{figure}
    \centering
    \includegraphics[width=0.55\linewidth]{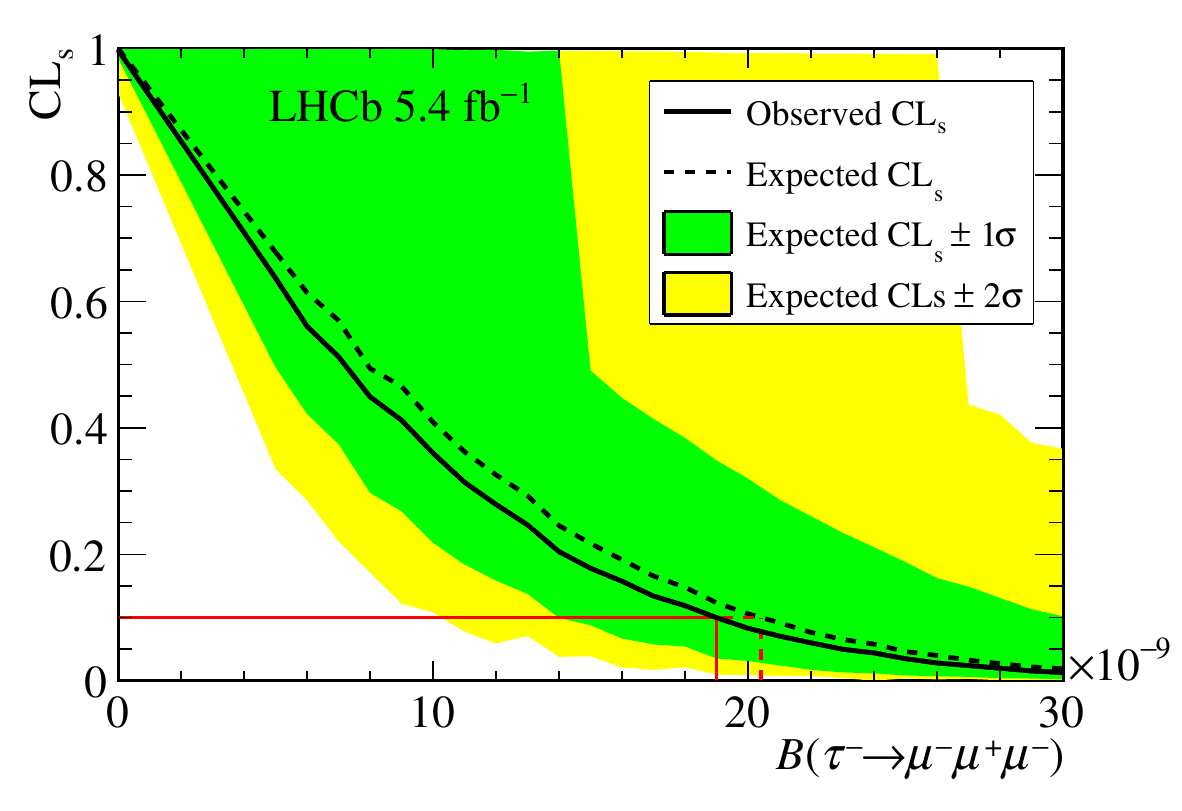}
    \caption{$\mathrm{CL}_s$ distribution as a function of the assumed branching fraction for \TauToMuMuMu decays, for both the observed data (solid line) and background-only expectation (dashed line) together with its intervals of one and two standard deviations ($\sigma$). The red line corresponds to the 90\% confidence level.}
    \label{fig:observed_limit}
\end{figure}

The central value of the signal branching fraction extracted from the fit is $\nolinebreak{(-0.1\pm1.1)\times 10^{-8}}$, where the uncertainty includes both statistical and systematic sources, leading to an observed upper limit on the branching fraction of
\[
\BF(\TauToMuMuMu)<1.9\,(2.3) \times 10^{-8}\,\text{at 90\% (95\%) CL.}
\]
The expected and observed $\text{CL}_\textit{s}$ distributions are shown in figure~\ref{fig:observed_limit} as a function of the assumed branching-fraction values. This result supersedes the LHCb Run~1 limit of \mbox{$4.6\,(5.6)\times10^{-8}$} at 90\% (95\%) CL~\cite{LHCb-PAPER-2014-052}.

\section{Conclusion}
\label{sec:Conclusion}

A search for the \TauToMuMuMu decay is carried out on data collected by the LHCb experiment between 2016 and 2018, corresponding to a total integrated luminosity of 5.4\invfb. No significant excess of signal is found and an upper limit of \mbox{$1.9\,(2.3) \times 10^{-8}$} at 90\% (95\%) CL is set on the branching fraction of the decay.
This result provides a complementary constraint on SM extensions in the context of existing searches, with a sensitivity comparable to that achieved by the \belletwo experiment \cite{Belle-II:2024sce}. 
Future analyses with the upgraded LHCb detector will benefit from higher luminosity and trigger efficiencies, leading to improved sensitivity to the signal decay.

\section*{Acknowledgements}
%
%
\noindent We express our gratitude to our colleagues in the CERN
accelerator departments for the excellent performance of the LHC. We
thank the technical and administrative staff at the LHCb
institutes.
We acknowledge support from CERN and from the national agencies:
ARC (Australia);
CAPES, CNPq, FAPERJ and FINEP (Brazil); 
MOST and NSFC (China); 
CNRS/IN2P3 (France); 
BMFTR, DFG and MPG (Germany);
INFN (Italy); 
NWO (Netherlands); 
MNiSW and NCN (Poland); 
MCID/IFA (Romania); 
MICIU and AEI (Spain);
SNSF and SER (Switzerland); 
NASU (Ukraine); 
STFC (United Kingdom); 
DOE NP and NSF (USA).
We acknowledge the computing resources that are provided by ARDC (Australia), 
CBPF (Brazil),
CERN, 
IHEP and LZU (China),
IN2P3 (France), 
KIT and DESY (Germany), 
INFN (Italy), 
SURF (Netherlands),
Polish WLCG (Poland),
IFIN-HH (Romania), 
PIC (Spain), CSCS (Switzerland), 
and GridPP (United Kingdom).
We are indebted to the communities behind the multiple open-source
software packages on which we depend.
Individual groups or members have received support from
Key Research Program of Frontier Sciences of CAS, CAS PIFI, CAS CCEPP, 
Minciencias (Colombia);
EPLANET, Marie Sk\l{}odowska-Curie Actions, ERC and NextGenerationEU (European Union);
A*MIDEX, ANR, IPhU and Labex P2IO, and R\'{e}gion Auvergne-Rh\^{o}ne-Alpes (France);
Alexander-von-Humboldt Foundation (Germany);
ICSC (Italy); 
Severo Ochoa and Mar\'ia de Maeztu Units of Excellence, GVA, XuntaGal, GENCAT, InTalent-Inditex and Prog.~Atracci\'on Talento CM (Spain);
SRC (Sweden);
the Leverhulme Trust, the Royal Society and UKRI (United Kingdom).




\addcontentsline{toc}{section}{References}
\bibliographystyle{LHCb/LHCb}
\bibliography{main,LHCb/standard,LHCb/LHCb-PAPER,LHCb/LHCb-CONF,LHCb/LHCb-DP,LHCb/LHCb-TDR}

\newpage
\centerline
{\large\bf LHCb collaboration}
\begin
{flushleft}
\small
R.~Aaij$^{38}$\lhcborcid{0000-0003-0533-1952},
A.S.W.~Abdelmotteleb$^{58}$\lhcborcid{0000-0001-7905-0542},
C.~Abellan~Beteta$^{52}$\lhcborcid{0009-0009-0869-6798},
F.~Abudin{\'e}n$^{58}$\lhcborcid{0000-0002-6737-3528},
T.~Ackernley$^{62}$\lhcborcid{0000-0002-5951-3498},
A. A. ~Adefisoye$^{70}$\lhcborcid{0000-0003-2448-1550},
B.~Adeva$^{48}$\lhcborcid{0000-0001-9756-3712},
M.~Adinolfi$^{56}$\lhcborcid{0000-0002-1326-1264},
P.~Adlarson$^{86}$\lhcborcid{0000-0001-6280-3851},
C.~Agapopoulou$^{14}$\lhcborcid{0000-0002-2368-0147}
% ...

{\footnotesize \it

% ...
$^{14}$Universit{\'e} Paris-Saclay, CNRS/IN2P3, IJCLab, Orsay, France\\
% ...
$^{38}$Nikhef National Institute for Subatomic Physics, Amsterdam, Netherlands\\
% ...
$^{48}$Instituto Galego de F{\'\i}sica de Altas Enerx{\'\i}as (IGFAE), Universidade de Santiago de Compostela, Santiago de Compostela, Spain\\
% ...
$^{52}$Physik-Institut, Universit{\"a}t Z{\"u}rich, Z{\"u}rich, Switzerland\\
% ...
$^{56}$H.H. Wills Physics Laboratory, University of Bristol, Bristol, United Kingdom\\
% ...
$^{58}$Department of Physics, University of Warwick, Coventry, United Kingdom\\
% ...
$^{61}$School of Physics and Astronomy, University of Glasgow, Glasgow, United Kingdom\\
$^{62}$Oliver Lodge Laboratory, University of Liverpool, Liverpool, United Kingdom\\
% ...
$^{70}$Syracuse University, Syracuse, NY, United States\\
% ...
$^{86}$Department of Physics and Astronomy, Uppsala University, Uppsala, Sweden, associated to $^{61}$\\
% ...
\medskip
}
\end{flushleft}

\end{document}